\documentclass[pre,twocolumn,showpacs,floatfix,superscriptaddress]{revtex4}
 \usepackage{graphicx}

 \newcommand{\be}{\begin{equation}}
 \newcommand{\ee}{\end{equation}}
 \newcommand{\bea}{\begin{eqnarray}}
 \newcommand{\eea}{\end{eqnarray}}
 \newcommand{\la}{\lambda}
 \newcommand{\xv}{\vec{x}}
 \newcommand{\ep}{\epsilon}
 \newcommand{\bav}{\langle b(t) \rangle}

\begin{document}

 \title{Survival probability of a diffusing test particle in a system of
coagulating and annihilating random walkers.}
 \author{R. Rajesh}
 \email{rrajesh@brandeis.edu}
 \affiliation{Martin Fisher School of Physics, Brandeis University,
Mailstop 057, Waltham, MA 02454-9110, USA}
 \author{Oleg Zaboronski}
 \email{olegz@maths.warwick.ac.uk}
 \affiliation{Mathematics Institute, University of Warwick, Gibbet Hill
Road, Coventry CV4 7AL, UK}


 \begin{abstract}
 We calculate the survival probability of a diffusing test particle in an
environment of diffusing particles that undergo
coagulation at rate $\lambda_c$ and annihilation at rate
$\lambda_a$. The test particle dies at rate $\lambda'$ on coming
into contact with the other particles. The survival probability
decays algebraically with time as $t^{-\theta}$. The exponent
$\theta$ in $d<2$ is calculated using the perturbative
renormalization group formalism as an expansion in $\epsilon=2-d$.
It is shown to be universal, independent of $\lambda'$, and to
depend only on $\delta$, the ratio of the diffusion constant of test
particles to that of the other particles, and on the ratio
$\lambda_a/\lambda_c$. In $2$-dimensions we calculate the
logarithmic corrections to the power law decay of the survival
probability. Surprisingly, the log corrections are non-universal.
The $1$-loop answer for $\theta$ in one dimension obtained by
setting $\epsilon=1$ is compared with existing exact solutions for
special values of $\delta$ and $\lambda_a/\lambda_c$. 
The analytical results for the
logarithmic corrections are verified by Monte Carlo simulations.
 \end{abstract}
 \maketitle

\section{\label{sec1}Introduction}

The calculation of the survival probability of a test particle in
reaction diffusion systems has been studied in different contexts
such as site persistence \cite{derrida1995}, walker persistence
problems
\cite{howard1996,monthus1996,majumdar1998,donoghue2002b,KRZ2003},
polydispersity exponents in models of aggregation
\cite{cueille1997,alava2002,KRZ2002,KRZ2003} and predator-prey
models \cite{redner1999,cardy2003}. The approach to these problems
have mostly been based on studying exactly solvable limiting cases
or using mean field approximation and its improvements such as the
Smoluchowski approximation
\cite{smoluchowski1917b,chandrasekhar1943}. In recent years, field
theoretic methods have proved successful in providing a general
framework to understand these problems. In particular,
the renormalization group analysis has been instrumental both in
identifying the universal persistent features of
reaction-diffusion systems and in extracting quantitative results
about persistence exponents which could not be obtained using
other methods
\cite{cardy2003,peliti1986,lee1994,cardy1995,howard1996,KRZ2002,KRZ2003}.
In this paper, we apply field theoretic methods to the problem of
the survival probability of diffusing test $B$-particles in a
background of diffusing $A$-particles undergoing the reactions
 \bea
 A + A & \stackrel{\lambda_c}{\longrightarrow} A, \nonumber\\
 A + A & \stackrel{\lambda_a}{\longrightarrow} \emptyset,
\label{eq:reaction}\\
 A + B & \stackrel{\lambda'}{\longrightarrow} A. \nonumber
 \eea

The above reaction has been studied in the context of persistence.
In one dimension and when $B$-particles are stationary,
calculating the survival probability of $B$-particles is
equivalent to calculating the fraction of spins that have not
flipped up to time $t$ in the $q$-state Potts model evolving via
zero temperature Glauber dynamics, where $q=\frac{\la_c}{\la_a}+2$
\cite{derrida1995}.  The more general problem in which the
$B$-particles are mobile with a diffusion constant equal to
$\delta$ times the diffusion constant of the $A$-particles has
been studied in
Refs.~\cite{howard1996,KNR1994,monthus1996,KRZ2003}. The density
of $B$-particles then decays with time as $t^{-\theta(\delta,Q)}$,
where $Q=(\lambda_c+\lambda_a)/ (\lambda_c+ 2 \lambda_a)$. As $Q$
varies from $1/2$ to $1$, the ratio $\lambda_c/\lambda_a$ varies
from $0$ to $\infty$. The known results for $\theta(\delta,Q)$ are
briefly reviewed below.

When the dimension $d$ is greater than the upper critical dimension- two
in this case- the decay exponents are obtained by solving the mean field
rate equations with appropriately renormalized lattice-dependent reaction
rate. In dimensions $d\leq 2$, fluctuation effects become important, and
$\theta(\delta,Q)$ is no longer given by the rate equations. Exact
solutions are one way of calculating exponents in $1$-dimension. When
$\delta=0$, by mapping the calculation of the persistence probability to a
calculation of empty interval probabilities in the $A+A \rightarrow A$
model, it was shown that \cite{derrida1995}
 \be
 \theta(0,Q)= \frac{2}{\pi^2} \left[\cos^{-1}\left( \frac{1 - 2 Q
}{\sqrt{2}} \right) \right]^2 -\frac{1}{8}, \quad d=1.
 \label{delta0}
 \ee
 Attempts to generalize the methods used in Ref.~\cite{derrida1995} to
arbitrary $\delta$ were successful only in determining the values of
$\frac{d \theta}{d Q}|_{Q=0}$ and $\frac{d \theta}{d \delta}|_{\delta=0}$
\cite{monthus1996}.  Another solvable limit is $Q=1$, when annihilation is
absent. In this case, the problem reduces to a three particle problem
which can be solved exactly to yield \cite{FG1988}
 \be
 \theta(\delta, 1)= \frac{\pi}{2 \cos^{-1}\left[ \delta/(1+\delta)
\right]}, \quad d=1.
 \label{Q1}
 \ee
 More general ways of understanding the effects of fluctuation in low
dimensional reaction-diffusion systems include the Smoluchowski
approximation \cite{smoluchowski1917b,chandrasekhar1943} which
effectively reduces to the replacement of the reaction rates in
the rate equations by  diffusion-renormalized values, and the
renormalization group formalism. The exponent $\theta(\delta,1/2)$
was calculated using the Smoluchowski approximation in
Ref.~\cite{KNR1994}. The advantage of Smoluchowski approximation
is its computational simplicity. However, it is not clear how to
improve this approximation in a systematic manner. Also, it was
shown in Refs.~\cite{KRZ2002,KRZ2003} that while this approach
gives an answer close to the actual one for $Q=1/2$, it becomes
increasingly worse as $Q$ nears $1$. The field theoretic approach
using the renormalization group formalism currently provides the
only systematic way of calculating the decay exponents below the
critical dimension. The exponent $\theta(\delta,1/2)$ was
calculated using field theoretic methods in Ref.~\cite{howard1996}. However,
the renormalization group scheme used was complicated and did not capture the
right logarithmic corrections (see Sec.~\ref{sec4b} for a more detailed
discussion). $\theta(1,Q)$ was
calculated as an expansion in $(2-d)$ in \cite{KRZ2003,KRZ2002} 
in the context of domain wall persistence in the Potts model.

In this paper, we extend the formalism developed in
Refs.~\cite{KRZ2003,KRZ2002}
to calculate $\theta(\delta, Q)$ for arbitrary $\delta$ and $Q$ to order
$\epsilon$, where $\epsilon=2-d$. In particular we show that
 \bea
 \lefteqn{\theta= \frac{Q (1+\delta)}{2} \times } \nonumber \\ && \left[ 2
- \epsilon \left\{ \frac{3}{2} + \ln \frac{1+\delta}{2} + \frac{Q
(1+\delta) f(\delta)}{2} \right\} + O(\epsilon^2) \right],
\label{four}
 \eea
 where
 \be
 f(\delta) = 1 - 2 \delta + 2 \delta \ln\left(\frac{2}{1+\delta} \right) +
(1-\delta^2) \int_{\frac{\delta-1}{\delta+1}}^{1} du \frac{\ln(1-u)}{u}.
 \label{eq:f}
 \ee
 The function $f(\delta)$ increases from $(1-\pi^2/4)$ to $0$ as $\delta$
increases from $0$ to $\infty$.  In $2$-dimensions, we calculate
logarithmic corrections to the power law decay and show that
 \be
 \langle b \rangle \sim t^{-Q(1+\delta)}\ln(t)^{\alpha/2}, \label{six}
 \ee
 where $\langle b \rangle$ is the mean density of $B$-particles and
 \bea
 \alpha &=& Q (1+\delta) \left[3 + Q (1+\delta)
f(\delta)+2 \ln \frac{1+\delta}{2} \right]  \nonumber \\
&&\mbox{}+ 4\pi Q(1+\delta)^2 \bigg(
\frac{1}{\lambda'}-\frac{2}{(1+\delta) (\lambda_a + \lambda_c)} \bigg),
\label{seven}
 \eea
 with the function $f$ as defined in Eq.~(\ref{eq:f}). A 
surprising feature of Eq.~(\ref{seven}) is its non-universality for
finite reaction rates, $\lambda, \lambda'<\infty$. In this case
$\alpha$ explicitly depends on both reaction rates. This is contrary to the
usual belief that below the upper critical dimension the kinetics is
diffusion limited and hence one may set reaction rates to infinite. Most
exact solutions make use of this simplifying assumption. The above result
serves as a counter example.

The rest of the paper is organized as follows. In Sec.~\ref{sec2},
the model is defined. In Sec.~\ref{sec3}, the rate equation
approach is compared with the Smoluchowski approximation. The
survival probability is calculated to one loop precision. In
Sec.~\ref{sec4}, the renormalization group analysis of the problem
is carried out and Eqs.~(\ref{four}), (\ref{six}), and 
(\ref{seven}) are derived. The one-loop answer for $\theta$ is
compared with the result of Smoluchowski approximation and also
with known exact results in $1$-dimension for special values of
$\delta$ and $Q$. We also compare the analytical
results with the results from  numerical simulations. First, the
predictions for the logarithmic corrections to the
power law decay are confirmed numerically in the limit of
instantaneous reactions. Second, the non-universality of logarithmic
corrections for finite reaction rates is verified.
Finally, we end with a summary and conclusions in Sec.~\ref{sec5}.

\section{\label{sec2}The model.}

Consider a $d$-dimensional hyper-cubic lattice whose sites may be occupied
by $A$-particles and $B$-particles. Multiple occupancy of a site is
allowed. Given a configuration of particles, the system evolves in time as
follows. (i) At rate $D/(2d)$, an $A$-particle hops to a nearest neighbor
site. (ii) At rate $\delta D/(2 d)$, a $B$-particle hops to a nearest
neighbor site.  (iii) At rate $ \lambda_a$, two $A$-particles at the same
site annihilate each other. (iv) At rate $\lambda_c$, two $A$-particles at
the same site coagulate together, thus reducing the number of $A$
particles by one. (v) At rate $\lambda'$, a $B$-particle is absorbed by an
$A$-particle at the same site. The initial number of $A$- ($B$-) particles
at the lattice sites are chosen independently from a Poisson distribution
with intensity $a_{0}$ ($b_{0}$).

The action corresponding to the continuous limit of the model can be
derived from the master equation using Doi's formalism
\cite{doi1976a,doi1976b,zeldovich1978}.
We skip the derivation and give the final result. The action is
 \bea
 S & = & \int dt \int d^d x \bigg( \bar{a} (\partial_t - \nabla^2) a
+\bar{b} (\partial_t - \delta \nabla^2) b + \frac{\lambda}{2 Q} \bar{a}
a^2 \nonumber \\ && \mbox{} + \frac{\lambda}{2} \bar{a}^2 a^2 + \lambda'
\bar{b} a b + \lambda' \bar{a} \bar{b} a b
-(\bar{a}a_{0}+\bar{b}b_{0})\delta(t) \bigg) \label{action}
 \eea
 where the $a$ and $b$ are complex fields corresponding to $A$- and
$B$-particles, the diffusion constant $D$ has been set equal to $1$, and
 \bea
 Q &=& \frac{\lambda_c+\lambda_a}{\lambda_c+2\lambda_a},\\
 \lambda&=& \lambda_a+\lambda_c.
 \eea
 The knowledge of all the correlation functions of the fields $a,b$ allows
one to construct all the correlation functions of local densities of $A$
and $B$ particles \cite{cardyhttp}. In particular, the mean density of $A$ and
$B$ particles at $(\xv ,t)$ is equal to $\langle a(\xv,t) \rangle$ and
$\langle b(\xv,t) \rangle$ respectively, where $\langle \ldots \rangle$
denotes functional average with respect to the functional measure
Eq.~(\ref{action}).

The action can be brought into a more convenient form by rescaling the
fields as follows: $(\bar{a}, \bar{b}) \rightarrow Q^{-1} (\bar{a},
\bar{b})$, $(a_{0}, b_{0}) \rightarrow Q (a_{0}, b_{0})$, and $(\lambda,
\lambda') \rightarrow 2 Q (\lambda, \lambda')$. Then
 \bea
 S & = & \int dt \int d^d x \bigg( \bar{a} (\partial_t - \nabla^2) a
+\bar{b} (\partial_t - \delta \nabla^2) b + \lambda \bar{a} a^2 \nonumber
\\ &+&
 \lambda \bar{a}^2 a^2 + 2 \lambda' Q \bar{b} a b + 2 \lambda' \bar{a}
\bar{b} a b -(\bar{a}a_{0}+\bar{b}b_{0})\delta(t) \!\bigg).
 \label{eq:action}
 \eea
 The Feynman diagrams corresponding to the action in Eq.~(\ref{eq:action})
are shown in Fig.~\ref{fig1}.

We are interested in the mean density of $B$-particles in the
limit of large time, as the survival probability is proportional
to mean density.
The evolution of mean density of $A$-particles $\langle a \rangle$
is independent of the statistics of $B$-particles and decays at
large times $t$ as \cite{peliti1986}
 \be
 \langle a \rangle \sim \cases{
 t^{-d/2}, &$d<2$, \cr
 t^{-1} \ln(t), &$d=2$, \cr
 t^{-1}, &$d>2$. \cr }
 \ee

\section{\label{sec3} The computation of persistence exponent using mean field and Smoluchowski approximations.}

\begin{figure}
 \includegraphics[width=8.0cm]{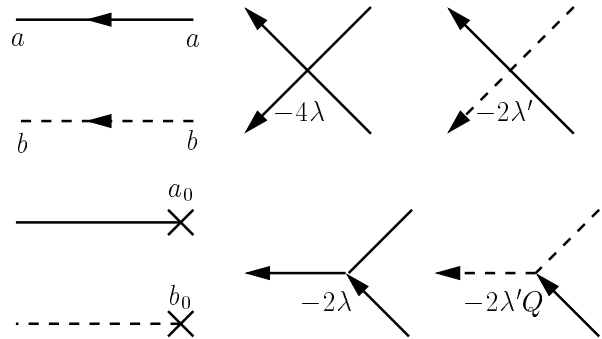}
 \caption{\label{fig1} Propagators and vertices of the theory.}
 \end{figure}

The perturbative expansion of $\langle b \rangle$ in powers of
$\lambda$ can be constructed using the Feynman diagrams shown in
Fig.~\ref{fig1} \cite{drouffe1994}. Diagrammatically, $\langle a
\rangle$ ($\langle b \rangle$) is the sum of all Feynman diagrams
with one outgoing $a$-($b$-)line respectively. Clearly, there is
an infinite number of diagrams contributing to $\langle a \rangle$
and $\langle b \rangle$. These diagrams can be grouped together
according to the number of loops that they contain, thus giving
rise to the loop expansion. Let $\epsilon = 2-d $. A simple
combinatorial argument shows that the contribution from a diagram
with $n$ loops is proportional to $g(t)^n$, where $g(t) = \lambda
t^{\epsilon/2}$ \cite{lee1994}. When $\epsilon <0$, the main
contribution to $\langle a \rangle$ and $\langle b \rangle$ comes
from properly renormalized tree level diagrams (diagrams without
loops). When $\epsilon>0$, the loop expansion fails since for
large times $g(t)$ is no longer a small perturbation parameter. We
therefore conclude that $2$ is the upper critical dimension. For
$d<2$ we will use the formalism of perturbative renormalization
group to convert the loop expansion into an $\epsilon$-expansion
and calculate scaling exponents as a series in $\epsilon$.

\subsection{\label{sec3a}Tree level diagrams}

\begin{figure}
 \includegraphics[width=8.0cm]{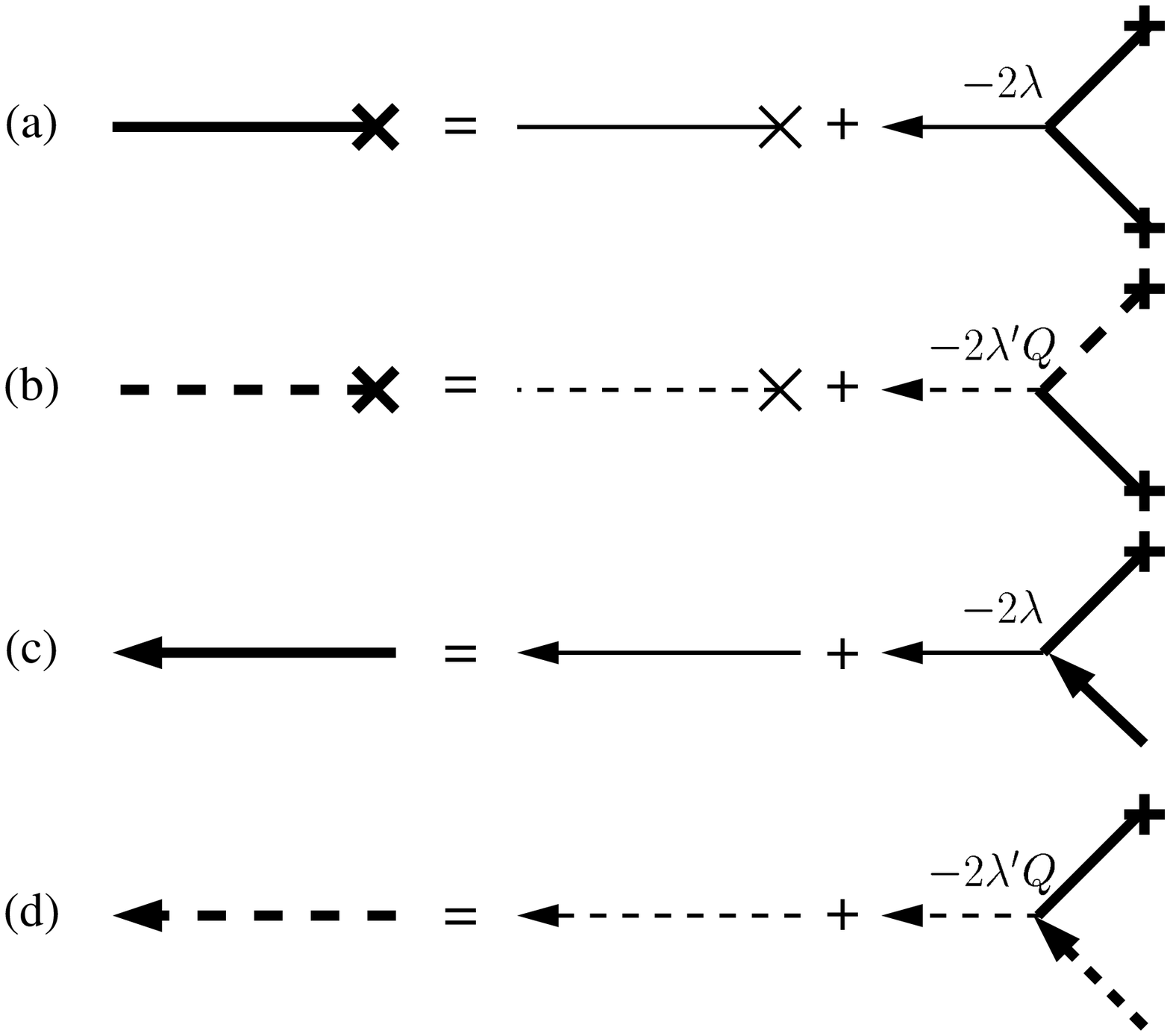}
 \caption{\label{fig2} Diagrammatic form of mean field equations for (a)
mean particle density $\langle b \rangle$, (b) mean density of
$B$-particles $\langle b \rangle$, (c) $G_{\rm{mf}}^{\rm{NN}}$, and (d)
$G_{\rm{mf}}^{\rm{PP}}$.}
 \end{figure}

In $d<2$ and small times, tree diagrams give the main contribution
to the survival probability. Let $\langle a \rangle_{\rm{mf}}$ and
$\langle b \rangle_{\rm{mf}}$ be mean field densities given by the
sum of contributions coming from tree diagrams with a single
outgoing $a$-line and $b$-line respectively.  We denote $\langle a
\rangle_{\rm{mf}}$ and $\langle b \rangle_{\rm{mf}}$ by thick
solid lines and thick dashed lines respectively. The integral
equations satisfied by $\langle a \rangle_{\rm{mf}}$ and $\langle
b \rangle_{\rm{mf}}$ are presented in diagrammatic form in
Fig.~\ref{fig2}(a)  and \ref{fig2}(b). After differentiating with
respect to time, they can be written in analytic form as
 \bea
 \partial_{t} \langle a \rangle &=& -\lambda \langle a \rangle^2
\label{Nmfeq}, \\
 \partial_{t} \langle b \rangle &=& -2 Q \lambda' \langle b \rangle
\langle a \rangle, \label{Pmfeq}
 \eea
 in which one can easily recognize the rate equations of the
 model. Thus, the identification of tree-level truncation with mean
 field approximation is justified.

Equations~(\ref{Nmfeq}) and (\ref{Pmfeq}) are easily solved yielding
 \bea
 \langle a(t) \rangle_{\rm{mf}} &=& \frac{a_{0}}{1+\lambda a_{0} t},
\label{Nmf} \\
 \langle b(t) \rangle_{\rm{mf}} &=& \frac{b_{0}}{(1+\lambda a_{0}t)^{2Q
\lambda'/\lambda}}, \label{Pmf}
 \eea
 where $a_0$ and $b_0$ are the initial densities of $A$- and $B$-particles
respectively. Thus,
 \bea
 \theta(\delta, Q)=2Q\frac{\la'}{\la}, \quad d>2. 
 \label{thmf}
 \eea
 The result is explicitly dependent on $\lambda', ~\lambda$ 
while being independent
of $\delta$ and describes the reaction-limited regime of the
problem. 
It should be mentioned here that the above
result is valid only in the limit when the reaction rates are the smallest
parameters in the problem, i.e. $\lambda,~\lambda' \ll l_0^{d-2}$, where $l_0$
is the lattice spacing. In the other limit when the lattice spacing is the
smallest parameter in the problem, the exponents gets modified to
\cite{RZunpublished}
 \be
 \theta(\delta, Q)=Q (1+\delta),\quad l_0^{d-2} \ll \lambda,~\lambda',\quad
d>2.
 \ee

In order to estimate the validity of the mean field  approximation
in $d\leq 2$, the one-loop corrections to the mean field answer have to be
evaluated. 
In calculating loop corrections to Eqs.~(\ref{Nmf}) and (\ref{Pmf}), we
are faced with the problem of summing over infinite number of diagrams
containing a given number of loops.  This problem can be simplified by
introducing mean field propagators which are sums of all tree diagrams
with one incoming line and one outgoing line. Expressed in terms of these
mean field propagators, there are only a finite number of diagrams with a
fixed number of loops.

Let $G_{\rm{mf}}^{\rm{NN}}$ and $G_{\rm{mf}}^{\rm{PP}}$ be mean field
propagators.  The integral equations satisfied by them are presented in
diagrammatic form in Figs.~\ref{fig2}(c)  and \ref{fig2}(d).  The
solutions to these equations are
 \bea
 G_{\rm{mf}}^{\rm{NN}}({\bf 2}|{\bf 1}) &= &\left( \frac{\langle a(t_2)
\rangle_{\rm{mf}})}{\langle a(t_1) \rangle_{\rm{mf}})} \right)^2
G_{1}({\bf 2}|{\bf 1}), \label{GNN} \\
 G_{\rm{mf}}^{\rm{PP}}({\bf 2}|{\bf 1}) & =& \left( \frac{\langle a(t_2)
\rangle_{\rm{mf}}}{\langle a(t_1) \rangle_{\rm{mf}}} \right)^{2 Q
\lambda'/\lambda} G_{\delta}({\bf 2}|{\bf 1}), \label{GPP}
 \eea
 where ${\bf 1}=(\vec{x}_{1},t_{1})$, ${\bf 2}=(\vec{x}_{2},t_{2})$ and
$G_{D}$ is the Green's function of the linear diffusion equation
with diffusion constant $D$.

\subsection{\label{sec3b}Smoluchowski approximation}

Before presenting the renormalization group calculation of $\theta
(\delta, Q)$, we briefly discuss a method commonly used to study
fluctuation effects in reaction-diffusion systems, namely the Smoluchowski
approximation. The essential idea of the Smoluchowski approach is to
relate the reaction rates $\lambda$ and $\lambda'$ to the diffusion rates.
One assumes that particles react instantaneously when they come within a
fixed radius of each other (see \cite{KNR1994,howard1996} for a more detailed
discussion). Knowing the first return probabilities of random walks, one
obtains for $d<2$,
 \bea
 \lambda &\sim& \rm{const} \times t^{d/2-1}, \\
 \lambda' &\sim& \rm{const} \times \left(\frac{1+\delta}{2}\right)^{d/2}
t^{d/2-1}.
 \eea
 In $d=2$ additional log corrections appear and
 \bea
 \lambda &\sim& \frac{\rm{const}}{\ln(t)}, \\
 \lambda' &\sim& \frac{\rm{const \times (1+\delta)}}{2
\ln[\left(\frac{1+\delta}{2} \right) t]}.
 \eea

Replacing $\lambda$ and $\lambda'$ in Eqs.~(\ref{Nmfeq}) and (\ref{Pmfeq})
by the effective reaction reaction rates, and solving for $\langle b
\rangle$, we obtain
 \be
 \langle b \rangle_{S} \sim \cases{ t^{- d Q (\frac{1+\delta}{2})^{d/2}},
&$d<2$, \cr \left(\frac{\ln(t)}{t}\right)^{Q (1+\delta)} (\ln(t))^{Q
(1+\delta)  \ln[(1+\delta)/2]}, &$d=2$. \cr} \label{eq:smol}
 \ee
 where $\langle b \rangle_S$ denotes the mean density of $B$-particles
obtained from Smoluchowski approximation. In particular, the Smoluchowski
theory's prediction for $\theta$ is
 \bea
 \theta_{S}(\delta,Q)= d Q \bigg( \frac{1+\delta}{2} \bigg)^{d/2}, ~d<2.
 \label{thsm}
 \eea
 This answer for $\theta$ depends on $\delta$, $Q$ and space
dimensionality $d$. It does not however depend on $\la$ and $\la'$. Thus,
unlike the mean field answer Eq.~(\ref{thmf}), it has the correct
universality properties for a quantity describing reaction-diffusion
systems in the diffusion-limited regime. However, the Smoluchowski result
differs considerably from the correct result when $Q$ nears $1$. For
example, in one dimension
$\theta_{S} (1,1)=1.0$ while $\theta(1,1)=1.5$ (see
Eq.~(\ref{Q1})). For more comparisons, see Sec.~\ref{sec4a}. It is not
clear how one can improve the Smoluchowski approximation.
The renormalization group formalism, though more involved, provides a
systematic way to go beyond the Smoluchowski approximation.

\subsection{\label{sec3c}One loop diagrams}

\begin{figure}
 \includegraphics[width=8.0cm]{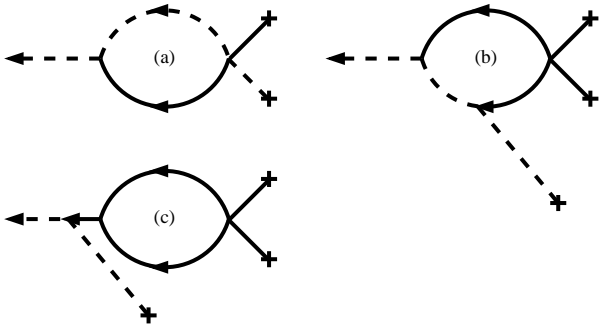}
 \caption{\label{fig3} One loop corrections to the mean field result for
$\langle b \rangle$.}
 \end{figure}

The rate equation results do not depend on the diffusion coefficients of
the particles or the dimensionality of the ambient space. These
parameters appear in the one-loop corrections to the tree level
answers. Using the mean field propagators and densities, it is
easy to classify all the one loop diagrams contributing to $\langle b
\rangle$. These are shown in Fig.~\ref{fig3}. Skipping the
computations we present the final answers for contributions
corresponding to each of the Feynman diagrams in the limit $a_{0}
\rightarrow \infty$:
 \bea
 (a) &=& \frac{ 32 Q \lambda'^2 b_0 t^{\epsilon/2}}
{\lambda 
(a_0 \lambda t)^{2 Q \lambda'/\lambda} 
[4 \pi (1+\delta)]^{d/2} 
\epsilon^2 (\epsilon+2)}, \\
 (b) &=& \frac{ 8 Q^2 \lambda'^2 b_0 (1+\delta) t^{\epsilon/2}} 
{\lambda
(a_0 \lambda t)^{2 Q \lambda'/\lambda} 
[4 \pi (1+\delta)]^{d/2} 
\epsilon }
\left[f(\delta) + O(\epsilon) \right], \\
 (c) &=& \frac{ -256 Q \lambda' b_0 t^{\epsilon/2}}{
(a_0 \lambda t)^{2 Q \lambda'/\lambda} 
(8 \pi)^{d/2} 
\epsilon^2 (\epsilon+2)^2 (\epsilon+4)},
 \eea
 where $(a)$, $(b)$ and $(c)$ refer to the contributions from diagrams in
Figs.~\ref{fig3}(a), \ref{fig3}(b) and \ref{fig3}(c)  respectively, and
 \be
 f(\delta) = 1 - 2 \delta + 2 \delta \ln\left(\frac{2}{1+\delta} \right) +
(1-\delta^2) \int_{\frac{\delta-1}{\delta+1}}^{1} du \frac{\ln(1-u)}{u}.
\label{fdef}
 \ee
 Adding these one-loop contributions to the mean field answer
Eq.~(\ref{Pmf}), we obtain in the limit $a_0 \rightarrow \infty$,
 \bea
 \lefteqn{ \langle b(t) \rangle= \frac{A}{t^{2 Q \lambda'/\lambda}}
\bigg[1 + \frac{8 Q \lambda' t^{\epsilon/2}} {(4 \pi)^{d/2} \epsilon }
\bigg\{ \frac{4 \lambda'}{\lambda (1+\delta)^{d/2}}
\frac{1}{\epsilon(\epsilon+2)} } \nonumber \\ &&\mbox{}-\frac{32}{2^{d/2}}
\frac{1}{\epsilon (\epsilon+2)^2 (\epsilon+4)} +\frac{Q \lambda'
(1+\delta)^{\epsilon/2} f(\delta)}{2 \lambda} \bigg\} \bigg] \nonumber \\
 && \mbox{} + \mbox{$2$- and higher loop corrections,} \label{oneloop}
 \eea
where $A=b_0/(a_0 \lambda)^{2 Q \lambda'/\lambda}$. We see that
if $\lambda \sim \lambda'$, then the  mean field answer Eq.~(\ref{thmf})
is correct in $d<2$ only if $Q \lambda t^{\ep/2} \ll 1$. Clearly, 
this condition breaks down in the limit of large times in $d<2$. 

\section{\label{sec4}Perturbative computation of $\theta(\delta,Q)$ near
$d=2$ using RG method.}

In this section, we calculate the large time behavior of $\langle
b \rangle$. The loop expansion for $\bav$ fails at large times in
$d\leq 2$. To extract the large time behavior of $\bav$ in $d \leq
2$ we will use the formalism of perturbative renormalization
group.

The renormalization group formalism used in
Refs.~\cite{KRZ2002,KRZ2003} for the case $\delta=1$ was based on
the Callan-Symanzik equations for the mean density of
$B$-particles. There were two relevant couplings for the theory:
the reaction rate $\lambda$ and the initial density $b_0$. The
anomalous dimension of $\bav$ was attributed to the
renormalization of $b_0$.

This approach turns out to be very complicated when $\delta\neq 1$. This is
due to the explicit
dependence of the classical scaling dimension of $\bav$
on $\lambda$ and $\lambda'$. Further complications arise due
to non-commutativity of $\epsilon \rightarrow 0$ and $a_{0}\rightarrow
\infty$ limits, which leads to an apparent order-$1/\ep^2$
singularity in the one-loop correction to $\bav$ [see Eq.~(\ref{oneloop})].

These problems are circumvented by analyzing the large time
asymptotic behavior of the logarithmic derivative of $\bav$,
rather than $\bav$ itself. It follows from
Eq.~(\ref{oneloop}) that
 \be
 t \partial_{t} [\ln(\bav)]=\Pi(t), \label{schd}
 \ee
 where
 \bea
 \Pi(t)= -2 Q \frac{\lambda'}{\lambda} + \frac{4 Q \lambda'
t^{\epsilon/2}} {(4 \pi)^{d/2} } \bigg\{ \frac{4 \lambda'}{\lambda
(1+\delta)^{d/2}}
\frac{1}{\epsilon(\epsilon+2)}  \nonumber \\
\mbox{} \!- \! \frac{2^{-d/2} 32}{\epsilon (\epsilon+2)^2
(\epsilon+4)} \! + \! \frac{Q \lambda' (1+\delta)^{\epsilon/2}
f(\delta)}{2 \lambda} \!\bigg\} \!\!  + \! O(\lambda^2).
\label{unreg}
\end{eqnarray}
 The large time asymptotic behavior of $\Pi (t)$ can be obtained by
solving the Callan-Symanzik equation with initial conditions given
by the right-hand side of Eq.~(\ref{unreg}) regularized at some
reference time $t_{0}$ (see Ref.~\cite{cardyhttp} for a review). The
coefficients of Callan-Symanzik equation are determined by the law
of renormalization of all the relevant couplings of the theory.
Power counting analogous to that carried out in Ref.~\cite{KRZ2002} shows
that there are only two relevant couplings of the theory which
determine large time behavior of $\Pi(t)$ in $d\leq 2$: the
reaction rates $\lambda$ and $\lambda'$. We mention here that $\Pi(t)$ is
simply related to the polarization operator used in Ref.~\cite{KRZ2002}.

Let $g_0= \lambda t_0^{\epsilon/2}$, $g'_0= \lambda
t_0^{\epsilon/2}$ be the dimensionless reaction rates. We choose
$t_{0}$ in such a way that $g_{0}$, $g'_{0} \ll 1$. The law of
renormalization of reaction rates has been worked out in
Ref.~\cite{peliti1986,howard1996}. The renormalized reaction rates $g_{R}$
and $g'_R$ are related to $g_0$ and $g'_0$ as follows:
 \bea
 g_{R} &=& \frac{g_0}{1+g_0/ g_*}, \label{gR} \\
 g'_{R} &=& \frac{g'_0}{1+g'_0/ g'_*}.
 \label{gRprime}
 \eea
 Here $g_*$ and $g'_*$ are the nontrivial fixed points of the
renormalization group flow in the space of effective coupling
constants, and are given by
 \bea
 g_* &=& \frac{(8 \pi)^{d/2}}{2 \Gamma(\epsilon/2)}, \\
 g'_* &=& \frac{[4 \pi (1+\delta)]^{d/2}}{2 \Gamma(\epsilon/2)},
 \eea
 where $\Gamma(x)$ is the Euler Gamma function.
The renormalization of both coupling constants is due to the same
physical effect -- the recurrence of random walks in $d\leq 2$.

$\Pi(t_{0})$ expressed in terms of $g_{R}$ and $g'_{R}$
has the following form:
 \bea
 \Pi(t_{0})&=&-2 Q
\frac{g'_{R}}{g_{R}} + \frac{ Q g'_{R}} { \pi } \bigg[ \frac{
g'_{R} (\gamma-1) } {g_{R} (1+\delta)}
+\frac{5-2\gamma}{4}
\nonumber\\
&&\mbox{}+ \frac{Q g'_{R}f(\delta)}{2 g_{R}} +O(\ep ) \bigg]
+ O(g_{R}^2), \label{reg}
 \eea
 where $\gamma$ is the Euler constant.
$\Pi (t_{0})$ regarded as a function of $g_{R}$ and
$g'_{R}$ is non-singular at $\ep=0$. As a result this expression
is valid for $d\leq 2$. The lack of $t_{0}$-dependence of $\Pi(t)$
for $t>t_{0}$ is expressed by the following renormalization group
(Callan-Symanzik) equation:
 \bea
  \bigg[ t \frac{\partial}{\partial t} +
 \frac{\beta(g_R)}{2} \frac{\partial}{\partial g_R}
+\frac{\beta(g'_R)}{2} \frac{\partial}{\partial g'_R}
   \big)
 \bigg] \Pi(t)=0, \label{CS2}
 \eea
 where $ \beta(g_R) = -2 t_0 \partial g_R/\partial t_0$ and
 $ \beta(g'_R) = -2 t_0 \partial g'_R/\partial t_0$ are the beta functions,
 \bea
 \beta(g_R) &=& \frac{g_R (g_R - g_*) \epsilon}{g_*}, \label{beta} \\
  \beta(g'_R) &=& \frac{g'_R (g'_R - g'_*) \epsilon}{g'_*}.\label{be2}
 \eea
 We will now solve Eq.~(\ref{CS2}) with the initial condition given by
Eq.~(\ref{reg})
to obtain the large time asymptotic behavior of $\Pi$. We then extract the
large time asymptotic behavior of $\bav$ by solving Eq.~(\ref{schd}).

\subsection{\label{sec4a}Survival probability in $d<2$}

At large times, the solutions of Callan-Symanzik
equation~(\ref{CS2}) are governed by the stable fixed points of
the beta functions. In $d<2$, these are $g_{R}=g_{*}$ and
$g'_{R}=g'_{*}$. It then follows that
 \bea
\Pi (t) &=& -2Q(1+\delta )
+ \ep Q(1+\delta)\bigg[ \ln\frac{1+\delta}{2}+\frac{3}{2}
\nonumber\\
&&\mbox{}+Q\frac{1+\delta}{2}f(\delta) \bigg]
+O(\ep^2, t^{-\ep/2}). \label{pi}
 \eea

Substituting Eq.~(\ref{pi}) into
Eq.~(\ref{schd}) and solving for $\bav$, we obtain the
$O(\epsilon)$ result for $\theta$:
 \bea
 \lefteqn{\theta= \frac{Q (1+\delta)}{2} \times } \nonumber \\ && \left[ 2
- \epsilon \left\{ \frac{3}{2} + \ln \frac{1+\delta}{2} + \frac{Q
(1+\delta) f(\delta)}{2} \right\} + O(\epsilon^2) \right],
 \label{thetaoneloop}
 \eea
 where the function $f(\delta)$ is as in Eq.~(\ref{fdef}).

\begin{figure}
 \includegraphics[width=\columnwidth]{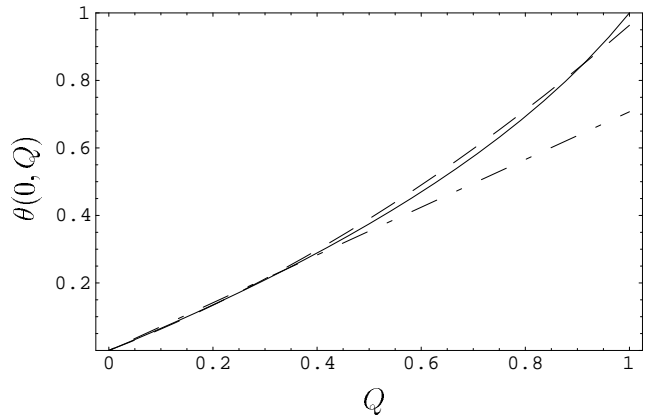}
 \caption{\label{fig4} The one-loop answer (Eq.~(\ref{thetaoneloop}))  is
compared with exact result in $1$-dimension when $\delta=0$
(Eq.~(\ref{delta0})). The solid line corresponds to the exact
answer, dashed line to one-loop, and dot dashed line to the
Smoluchowski result (Eq.~(\ref{eq:smol})).}
 \end{figure}

\begin{figure}
 \includegraphics[width=\columnwidth]{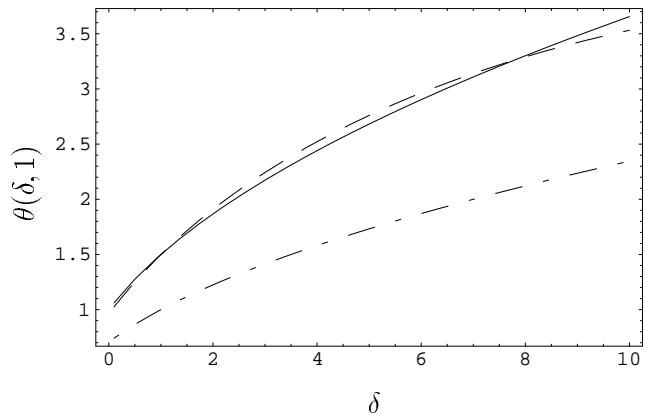}
 \caption{\label{fig5} The one-loop answer (Eq.~(\ref{thetaoneloop}))  is
compared with exact result in $1$-dimension when $Q=1$
(Eq.~(\ref{Q1})). The solid line corresponds to the exact answer,
dashed line to one-loop, and dot dashed line to the Smoluchowski
result (Eq.~(\ref{eq:smol})).}
 \end{figure}

We now compare the $1$-dimensional result obtained by putting
$\epsilon=1$ in Eq.~(\ref{thetaoneloop}) with exact results in
$1$-dimension for special values of $\delta$ and $Q$. The exact
result for $\theta(0, Q)$ in $1$-dimension is given in
Eq.~(\ref{delta0}), while that for $\theta(\delta,1)$ is given in
Eq.~(\ref{Q1}).  Figs~\ref{fig4} and \ref{fig5} show the results
for $\delta=0$ and $Q=1$ respectively. The $\epsilon$-expansion
result is seen to compare very well with the exact result. On the
other hand, the Smoluchowski results fail for $Q$ larger than
$1/2$ when $\delta=0$, and for all $\delta$ when $Q=1$. It should
be noted that when $\delta$ becomes large, the
$\epsilon$-expansion should fail and Smoluchowski result should
give a much better approximation.

\subsection{\label{sec4b}Survival probability in $d=2$.}

The upper critical dimension of our model is two. The non-trivial
fixed points of the $\beta$-functions Eqs.~(\ref{beta}) and
(\ref{be2}) vanish at $d=2$.  It is then expected that the rate equation
results give the correct large time behavior of mean
densities, perhaps modulo logarithmic corrections. This turns out
to be incorrect. The complication comes from the fact that
$\theta$ predicted by the mean field theory (see Eq.~(\ref{thmf}))  is
non-universal and depends on the ratio of coupling constants
$g_{R}$ and $g'_{R}$. Each of these couplings is marginally
relevant in two dimensions and flows under RG transformations to
$0$ as $[\ln(t)]^{-1}$. Their ratio flows to a finite
universal value which determines the algebraic decay of the
survival probability.  However, the deviation from this universal
value vanishes with time as $C/\ln(t)$, where C is a non-universal
constant. This results in non-universal logarithmic
corrections to the universal power law decay of survival
probability. It is worth mentioning that in
$d<2$, similar considerations lead to a conclusion that the
amplitude of corrections to scaling is non-universal.

We need to solve the Callan-Symanzik equation (\ref{CS2}) with
coefficients calculated at $d=2$:
 \bea
 \beta(g)|_{d=2} &=& \frac{g^2}{2 \pi},\\
 \beta(g')|_{d=2} &=& \frac{g'^2}{ \pi (1+\delta)}.
 \eea
 Then Eq.~(\ref{CS2}) reduces to
 \bea
 \bigg[ t \frac{\partial}{\partial t} + \frac{g_R^2}{4 \pi}
\frac{\partial}{\partial g_R} + \frac{g^{'2}_{R}}{2 \pi
(1+\delta)} \frac{\partial}{\partial g'_R}
 \bigg] \Pi(t) =0, \label{CS2d}
 \eea
 which has to be solved with the initial condition given by Eq.~(\ref{reg})
at $t=t_{0}$, $\ep=0$.
  The solution is
 \bea
&& \Pi(t)=-Q(1+\delta)+
\frac{2Q(1+\delta)}{\ln(t/t_{0})}
\bigg[\frac{3}{4}+\frac{Q}{2} f(\delta) \nonumber\\
&&\mbox{}
+\pi(1+\delta)\left(\frac{1}{g'_{R}}-\frac{2}{(1+\delta) g_{R}}  \right)
\bigg]+O\bigg( \frac{1}{\ln^{2}(t)}\bigg). \label{pi2d}
 \eea

The non-universal term in Eq.~(\ref{pi2d}) is proportional to
$1/g'_{R}-2/[(1+\delta) g_{R}]$. It is convenient to
express this amplitude in terms of bare
couplings. In two dimensions,
 \bea
\frac{1}{g'_{R}}-\frac{2}{(1+\delta) g_{R}}=\frac{1}{g'_{0}}-\frac{2}{(1+\delta)
g_{0}}+\frac{\ln(\frac{1+\delta}{2})}{2\pi (1+\delta)}. \label{inv}
 \eea
In $d<2$, Eq.~(\ref{inv}) has to be
modified by the omitting the logarithmic term on the right hand side.
 \begin{figure}
 \includegraphics[width=8.0cm]{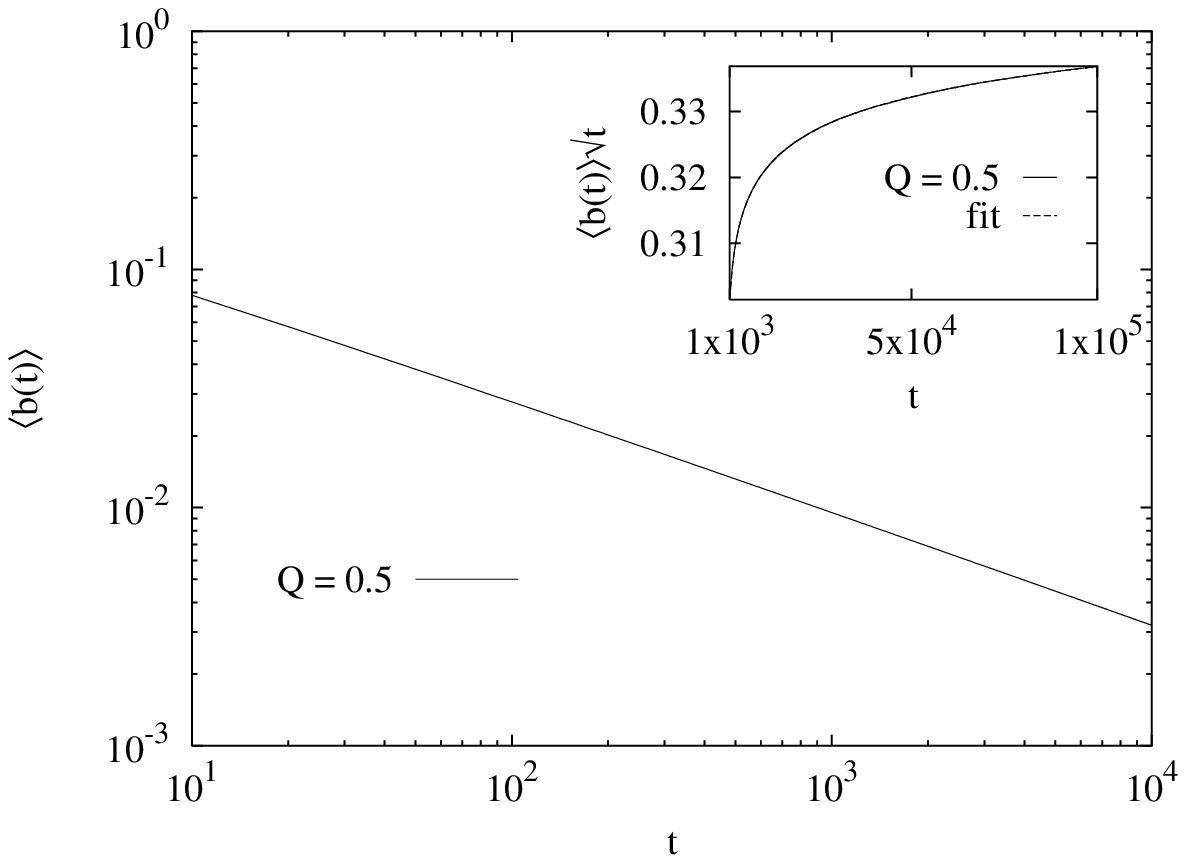}
 \caption{\label{fig6} The variation of the mean density of $B$-particles
in two dimensions with time is shown. The simulations were done on a
$3200\times 3200$ lattice for $Q=0.5$ and $\delta=0$. The data has been
averaged $1000$ times. In the inset, the variation of $\langle b \rangle
\sqrt{t}$ with time is shown. The power of the logarithm in the best fit is
$0.23\pm 0.03$.
}
 \end{figure}

Solving Eq.~(\ref{schd}) with Eq.~(\ref{pi2d}) substituted for the right hand
side and taking (\ref{inv}) into account, one finds that
 \be
 \langle b(t) \rangle = {\mbox{const}} \times \frac{\left(g_R
\ln(t/t_0)\right)^{\alpha/2}} {(g_R t)^{Q(1+\delta)}} \left(1+ O
(\frac{1}{\ln(t/t_0)}) \right), \label{pb2d}
 \ee
 where
 \bea
 \alpha &=& Q (1+\delta) \left[3 + Q (1+\delta)
f(\delta)+2 \ln \frac{1+\delta}{2} \right]  \nonumber \\
&&\mbox{}+ 4\pi Q(1+\delta)^2 \bigg( \frac{1}{\lambda'}-\frac{2}{(1+\delta)
\lambda} \bigg).
 \label{pb2dlog}
 \eea
Thus, in two dimensions, the power law exponent
is universal and
independent of $\lambda$ and $\lambda'$. However,
the logarithmic corrections are
non-universal and depend on microscopic reaction rates
$\lambda'=g'_{0}$ and $\lambda=g_{0}$. However, most simulations
are done in the limit when the reactions are instantaneous. In
this limit, the non-universal term in Eq.~(\ref{pb2dlog}) is zero.

The log corrections in Eq.~(\ref{pb2dlog}) are different from the
log corrections calculated for the $Q=1/2$ case in
Ref.~\cite{howard1996}. This discrepancy is due to the fact that only
renormalized tree level computations were done in \cite{howard1996},
while to obtain the correct logarithmic dependence, one-loop
corrections have to be taken into account.
 \begin{figure}
 \includegraphics[width=8.0cm]{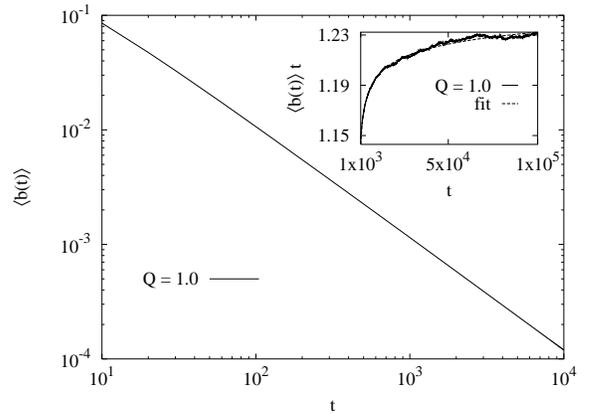}
 \caption{\label{fig7} The variation of the mean density of $B$-particles
in two dimensions with time is shown. The simulations were done on a
$3200\times 3200$ lattice for $Q=1.0$ and $\delta=0$. The data has been
averaged $1000$ times. In the inset, the variation of $\langle b \rangle
t$ with time is shown. The power of the logarithm in the best fit is
$0.08\pm 0.04$.
}
 \end{figure}
 \begin{figure}
 \includegraphics[width=8.0cm]{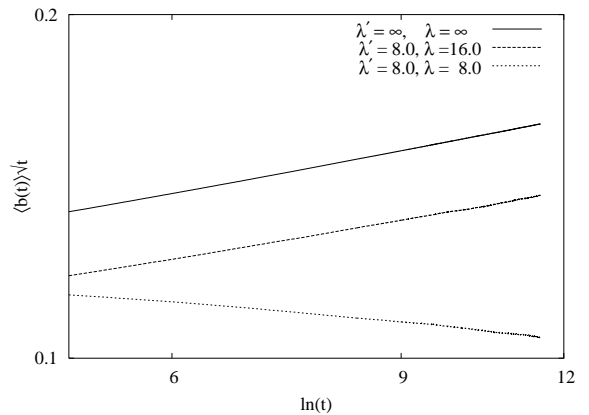}
 \caption{\label{fig8} The variation of the mean density of $B$-particles
in two dimensions with time is shown. The simulations were done on a
$900\times 900$ lattice for $Q=0.5$ and $\delta=0$.
The data has been averaged $1000$ times. }
 \end{figure}

We also mention that if one were to ignore the contribution from one-loop
diagrams, then the log corrections would be identical with the log
corrections obtained from the Smoluchowski approximation [see
Eq.~(\ref{eq:smol})],
and will be different from the log corrections obtained from renormalized
tree-level as in Ref.~\cite{howard1996}.

We now study logarithmic corrections numerically.
First, consider the case when the microscopic reactions are instantaneous,
i.e., $\lambda=\lambda'=\infty$. In this limit, the non-universal term in
Eq.~(\ref{pb2dlog}) is equal to zero. 
The Monte Carlo simulations were done
for this case on a two dimensional lattice of size $3200 \times
3200$ with periodic boundary conditions.  As the reactions are
instantaneous, the maximum number of particles at a site is one.
The simulations were done for $\delta=0$, i.e., immobile
$B$-particles. The results for $Q=0.5$ and $Q=1.0$ are shown in
Figs.~\ref{fig6} and \ref{fig7} respectively. For $Q=0.5$,
$\alpha= 0.23\pm 0.03$ which compares well with the theoretical
value of approximately $0.22$ in Eq.~(\ref{pb2dlog}). Note that
renormalized tree-level answer is $\alpha=0.5$ \cite{howard1996}.
For $Q=1.0$, the numerical value is $.08\pm 0.04$, which compares
well with the theoretical value of approximately $0.07$. This
answer also deviates significantly from renormalized tree-level
value of $1.0$.

Second, we study the logarithmic corrections for finite reaction
rates to test the non-universal term in Eq.~(\ref{pb2dlog}). In this
case the lattice size is $900\times 900$. The results for
$\delta=0$, $Q=0.5$ and different reaction rates are shown in
Fig.~\ref{fig8}. If $\lambda=16,~\lambda'=8$, then the
non-universal term in Eq.~(\ref{pb2dlog}) is zero and $\alpha \approx
0.22 $ as in the case of infinite reaction rates. This is
consistent with dashed line in Fig.~\ref{fig8} being parallel to
the solid line. If however, $\lambda=8,~\lambda'=8$, then $\alpha
\approx -0.57$. The slope of the bottom line in Fig.~\ref{fig8}
is clearly negative.

\section{\label{sec5}Summary and conclusions}

In summary, we calculated the large time behavior of the survival
probability of a test particle in a system of coagulating and
annihilating random walkers in $d\leq 2$. In one dimension, this
generalizes the site persistence problem in the $q$-state Potts
model evolving via zero temperature Glauber dynamics. The survival
probability was shown to decay as a power law. In $d<2$, the
exponent $\theta$ characterizing this power law was shown to be
universal, in the  sense that it depends only on $\delta$ and $Q$ and
was independent of $\lambda, ~\lambda'$. The renormalization group
formalism provided a systematic way of calculating the survival
probability for the entire parameter space.

In two dimensions, we computed the logarithmic corrections to the power law
decay. It was shown that to compute the correct logarithmic factors, one had
to include contributions from one-loop diagrams and not just the tree level
diagrams as was done in earlier work. 
The behavior of the survival probability in two dimensions is
surprising. First, the power law decay is universal and thus
does not coincide with the rate equation result, even though $d=2$ is
the upper critical dimension. 
Second, the logarithmic corrections to the power law are
non-universal and depend on the reaction rates. This is contrary to the
general expectation that kinetics of reaction-diffusion systems are diffusion
limited below the upper critical dimension. 
Both the universality of the
power law and the non-universality of log-corrections in two
dimensions can be traced to the fact that the rate equation
exponent is given by the ratio of microscopic rates,
which are both marginally relevant in two dimensions.

From the renormalization group point of view, by studying the logarithmic
derivative of the mean density of $B$-particles, we have considerably 
simplified the schemes used in Refs.~\cite{howard1996,KRZ2002,KRZ2003}. 
While the exponents are only calculated up to first order in $\epsilon$, the
renormalization group method remains the only systematic way of computing the
exponents when an exact solution is not available.

\section*{Acknowledgments}

RR would like to acknowledge support from NSF grant DMR-0207106.


\end{document}